\pdfoutput=1
\RequirePackage{ifpdf}
\ifpdf 
\documentclass[pdftex]{sigma}
\else
\documentclass{sigma}
\fi

\DeclareMathOperator{\diag}{diag}

\begin{document}

\newcommand{\arXivNumber}{1402.3541}

\allowdisplaybreaks

\renewcommand{\PaperNumber}{084}

\FirstPageHeading

\ShortArticleName{A Compact Formula for Rotations as Spin Matrix Polynomials}

\ArticleName{A Compact Formula for Rotations\\ as Spin Matrix Polynomials}

\Author{Thomas L.~CURTRIGHT~$^{\dag}$, David B.~FAIRLIE~$^\ddag$ and Cosmas K.~ZACHOS~$^{\S}$}

\AuthorNameForHeading{T.L.~Curtright, D.B.~Fairlie and C.K.~Zachos}

\Address{$^{\dag}~$Department of Physics, University of Miami, Coral Gables, FL 33124-8046, USA}
\EmailD{\href{mailto:curtright@miami.edu}{curtright@miami.edu}}
\URLaddressD{\url{http://www.physics.miami.edu/~curtright/}}

\Address{$^\ddag$~Department of Mathematical Sciences, Durham University, Durham, DH1 3LE, UK}
\EmailD{\href{mailto:David.Fairlie@durham.ac.uk}{David.Fairlie@durham.ac.uk}}

\Address{$^\S$~High Energy Physics Division, Argonne National Laboratory, Argonne, IL 60439-4815, USA}
\EmailD{\href{mailto: c.zachos.k@att.net}{c.zachos.k@att.net}}
\URLaddressD{\url{http://www.hep.anl.gov/czachos/}}

\ArticleDates{Received May 07, 2014, in f\/inal form August 07, 2014; Published online August 12, 2014}

\Abstract{Group elements of ${\rm SU}(2)$ are expressed in closed form as f\/inite polynomials of the Lie algebra generators,
for all def\/inite spin representations of the rotation group.
The simple explicit result exhibits connections between group theory, combinatorics, and Fourier analysis, especially in
the large spin limit.
Salient intuitive features of the formula are illustrated and discussed.}


\Classification{15A16; 15A30}

\section{Introduction}

Rotation matrices, worked out by Wigner~\cite{wigner} for all representations, are one of the mainstays of quantum
mechanics.
 They amount to group elements of ${\rm SU}(2)$, exponentials $e^{i\theta(\hat{\boldsymbol{n}}\cdot\boldsymbol{J})}$ of
suitable Lie algebra ($[J_{a},J_{b}]=i\epsilon_{abc}J_{c}$) generators.
 Of course, these are $(2j+1)\times(2j+1)$ dimensional matrices for either integral spin, $j=0,1,2,3,\dots$ or
half-integral spin, $j=\frac{1}{2}$, $\frac{3}{2}$, $\frac{5}{2}, \dots$.
 Here, $\theta$~is the rotation angle around the unit vector axis $\hat{\boldsymbol{n}}$, and the third component of
the matrix triplet~$\boldsymbol{J}$ in standard convention is the diagonal matrix
$J_{3}=\diag(j,j-1,j-2,\dots,1-j,-j)$.

By the Cayley--Hamilton theorem, the expansion of this exponential in powers of
$\hat{\boldsymbol{n}}\cdot\boldsymbol{J}$ may be recast as a~\emph{finite} sum of powers of
$\hat{\boldsymbol{n}}\cdot\boldsymbol{J}$, the highest power being of order~$2j$.
Such polynomials of Lie generators, in the universal enveloping algebra of~$su(2)$, have numerous celebrated
applications in physics.

Nevertheless, beyond standard expressions for $j=1/2$, so $\boldsymbol{J} =\boldsymbol{\sigma}/2$,
\begin{gather*}
e^{i(\theta/2)(\hat{\boldsymbol{n}}\cdot\boldsymbol{\sigma})}=I_{2}\cos{\theta/2}+i(\hat{\boldsymbol{n}}\cdot\boldsymbol{\sigma})\sin{\theta/2},
\end{gather*}
and the triplet, $j=1$, so $J_{3}=\diag(1,0,-1)$,
\begin{gather*}
e^{i\theta(\hat{\boldsymbol{n}}\cdot\boldsymbol{J})}=I_{3}+i(\boldsymbol{\hat
{n}}\cdot\boldsymbol{J})\sin{\theta}+(\hat{\boldsymbol{n}}\cdot\boldsymbol{J})^{2}(\cos\theta-1)
\\
\phantom{e^{i\theta(\hat{\boldsymbol{n}}\cdot\boldsymbol{J})}}
=I_{3}+(2i\hat{\boldsymbol{n}}\cdot\boldsymbol{J}\sin(\theta/2))\cos
(\theta/2)+\tfrac{1}{2}(2i\hat{\boldsymbol{n}}\cdot\boldsymbol{J}\sin (\theta/2))^{2},
\end{gather*}
which actually amounts to the vector Rodrigues' rotation formula~\cite{euler,rodrigues}, such expansions for higher
dimensionality multiplets are relatively rare.

While van Wageningen~\cite{wageningen} and Lehrer--Ilamed~\cite{lehrer} made substantial progress and provided some
signif\/icant results (also see the less direct algorithms in~\cite{WW} and~\cite{torruella}), until now there has not
appeared \emph{a simple, compact, closed expression} illuminating the properties of the expansion for arbitrary~$j$.
Our methods and spirit clearly overlap those of ref~\cite{wageningen}, although, crucially, our basic variable is
$\sin(\theta/2)$ and not trigonometric functions of $n\theta$ as in this and other treatments.
Because of this, we are able to furnish the {\em
ultimate solution} of all the algorithms appearing previously in the literature.

\section{New compact results}

A~formula valid for arbitrary spin~$j$ is given~by
\begin{gather}
e^{i \theta(\hat{\boldsymbol{n}}\cdot\boldsymbol{J})} = \sum\limits_{k=0}^{2j} \frac{c_k(\theta)}{k!}
\big(2i\hat{\boldsymbol{n}}\cdot\boldsymbol{J} \sin (\theta/2)\big)^k,
\label{sum}
\end{gather}
where, with use of the f\/loor function,
\begin{gather}
c_{k}(\theta)=(\cos(\theta/2))^{\epsilon}
\underset{\lfloor j-k/2\rfloor}{\operatorname{Trunc}}\left(\frac{(\arcsin\sqrt{x}/\sqrt{x})^{k}} {(\sqrt{1-x})^{\epsilon}}\right)
\label{trunco}
\end{gather}
with $x=\sin^{2}(\theta/2)$ and
\begin{gather*}
\epsilon(j-k/2)=2j-k-2\lfloor j-k/2\rfloor=\frac{1-(-1)^{2j-k}}{2},
\end{gather*}
a~binary ``parity''
variable: 0 for even $2j-k$, and 1 for odd $2j-k$.
The ef\/fective variable $2j-k$ is the descending order of this polynomial~\eqref{sum}.
Trunc$_{n}$ is a~Taylor polynomial in~$x$, i.e., it denotes truncating the inf\/inite series of its arguments to $O(x^{n})$, 
\begin{gather*}
{\operatorname*{Trunc}_{n}}\left(\sum\limits_{m=0}^{\infty}a_{m}x^{m}\right) \equiv\sum\limits_{m=0}^{n}a_{m}x^{m}.
\end{gather*}
Each term in the Taylor polynomials $\operatorname{Trunc}_n$ of~\eqref{trunco} is positive semi-def\/inite.

The compact formula~\eqref{sum} and its essential features are exemplif\/ied and analyzed in the sections to follow.
 Examples are given in Section~\ref{section3}.
 Reformulated in descending order, the polynomial coef\/f\/icients are discussed and inter-related in Section~\ref{section4}.
In Section~\ref{section5}, these coef\/f\/icients are organized into compact generating functions, linking dif\/ferent values of~$j$.
 The machinery needed to derive the results is presented in Section~\ref{section6}.
 Concluding commentary with special attention to the large spin limit is provided in Section~\ref{section7}, and illustrated in an
appendix.

\section{Basic examples and symmetries}\label{section3}

A~few low spin~$j$ cases of~\eqref{sum} are provided to exemplify the structure of the series.

For the quartet, $j=3/2$,
\begin{gather}
e^{i \theta (\hat{\boldsymbol{n}}\cdot\boldsymbol{J})} = I_4 \cos (\theta/2)\left(1+\tfrac{1}{2}\sin^2
(\theta/2)\right)+(2i \hat{\boldsymbol{n}}\cdot\boldsymbol{J} \sin (\theta/2))\left(1+\tfrac{1}{6} \sin^2 (\theta/2)
\right)
\nonumber
\\
\phantom{e^{i \theta (\hat{\boldsymbol{n}}\cdot\boldsymbol{J})}=}{}
+\frac{1}{2!} \bigl (2i \hat{\boldsymbol{n}}\cdot\boldsymbol{J}\sin (\theta/2) \bigr)^2 \cos (\theta/2)+\frac {1}{3!}
\bigl (2i \hat{\boldsymbol{n}}\cdot\boldsymbol{J}\sin (\theta/2) \bigr)^3.
\label{quartet}
\end{gather}

For the quintet, $j=2$,
\begin{gather*}
e^{i \theta(\hat{\boldsymbol{n}}\cdot\boldsymbol{J})} = I_5+(2i \hat{\boldsymbol{n}}\cdot\boldsymbol{J} \sin(\theta/2))
\cos(\theta/2)\left(1+\tfrac{2}{3}\sin^2(\theta/2)\right)
\\
\phantom{e^{i \theta (\hat{\boldsymbol{n}}\cdot\boldsymbol{J})}=}{}
+\frac{1}{2!} {(2i \hat{\boldsymbol{n}}\cdot\boldsymbol{J} \sin(\theta/2))^2}\left(1+\tfrac{1}{3} \sin^2 (\theta/2)\right)
\\
\phantom{e^{i \theta (\hat{\boldsymbol{n}}\cdot\boldsymbol{J})}=}{}
+\frac{1}{3!} {(2i \hat{\boldsymbol{n}}\cdot\boldsymbol{J} \sin(\theta/2))^3} \cos(\theta /2) +\frac{1}{4!} (2i
\hat{\boldsymbol{n}}\cdot\boldsymbol{J} \sin (\theta/2))^4.
\end{gather*}

For the sextet, $j=5/2$,
\begin{gather*}
e^{i \theta(\hat{\boldsymbol{n}}\cdot\boldsymbol{J})} = I_6 \cos(\theta/2)\left(1+ \tfrac{1}{2} \sin^2
(\theta/2+\tfrac{3}{8} \sin^4 (\theta/2)\right)
\\
\phantom{e^{i \theta(\hat{\boldsymbol{n}}\cdot\boldsymbol{J})} =}{}
+(2i \hat{\boldsymbol{n}}\cdot\boldsymbol{J} \sin(\theta /2))\left(1+\tfrac{1}{6}\sin^2(\theta/2)
+\tfrac{3}{40}\sin^4(\theta /2)\right)
\\
\phantom{e^{i \theta(\hat{\boldsymbol{n}}\cdot\boldsymbol{J})} =}{}
+\frac{1}{2!} {(2i \hat{\boldsymbol{n}}\cdot\boldsymbol{J} \sin(\theta/2))^2} \cos(\theta /2)
\left(1+\tfrac{5}{6}\sin^2(\theta/2)\right)
\\
\phantom{e^{i \theta(\hat{\boldsymbol{n}}\cdot\boldsymbol{J})} =}{}
+\frac{1}{3!} {(2i \hat{\boldsymbol{n}}\cdot\boldsymbol{J} \sin (\theta/2))^3}\left(1+\tfrac{1}{2}\sin^2(\theta/2)
\right)
\\
\phantom{e^{i \theta(\hat{\boldsymbol{n}}\cdot\boldsymbol{J})} =}{}
+\frac{1}{4!} {(2i \hat{\boldsymbol{n}}\cdot\boldsymbol{J} \sin(\theta/2))^4}\cos(\theta /2) +\frac{1}{5!} {(2i
\hat{\boldsymbol{n}}\cdot\boldsymbol{J} \sin (\theta/2))^5}.
\end{gather*}

For spin $j=5$,
\begin{gather*}
e^{i \theta(\hat{\boldsymbol{n}}\cdot\boldsymbol{J})} = I_{11}
+(2i \hat{\boldsymbol{n}}\cdot\boldsymbol{J} \sin(\theta/2)) \cos(\theta/2)
\\
\phantom{e^{i \theta(\hat{\boldsymbol{n}}\cdot\boldsymbol{J})} = }{}
\times\left(1+\tfrac{2}{3}\sin^2(\theta/2)
+\tfrac{8}{15}\sin^4(\theta/2) +\tfrac{16}{35}\sin^6(\theta/2) +\tfrac{128}{315}\sin^8(\theta/2)\right)
\\
\phantom{e^{i \theta(\hat{\boldsymbol{n}}\cdot\boldsymbol{J})} = }{}
+\tfrac{1}{2!} {(2i \hat{\boldsymbol{n}}\cdot\boldsymbol{J} \sin(\theta/2))^2}\\
\phantom{e^{i \theta(\hat{\boldsymbol{n}}\cdot\boldsymbol{J})} = }{}
\times
\left(1+\tfrac{1}{3} \sin^2 (\theta /2)
+\tfrac{8}{45} \sin^4 (\theta /2)+ \tfrac{4}{35} \sin^6 (\theta /2)+ \tfrac{128}{1575} \sin^8 (\theta /2)\right)
\\
\phantom{e^{i \theta(\hat{\boldsymbol{n}}\cdot\boldsymbol{J})} = }{}
+ \tfrac{1}{3!} {(2i \hat{\boldsymbol{n}}\cdot\boldsymbol{J} \sin(\theta/2))^3}\cos(\theta /2)\left(1+\sin^2 (\theta
/2)+ \tfrac{14}{15} \sin^4 (\theta /2)+ \tfrac{164}{189} \sin^6 (\theta /2)\right)
\\
\phantom{e^{i \theta(\hat{\boldsymbol{n}}\cdot\boldsymbol{J})} = }{}
+\tfrac{1}{4!} {(2i \hat{\boldsymbol{n}}\cdot\boldsymbol{J} \sin(\theta/2))^4}\left(1+\tfrac{2}{3} \sin^2 (\theta /2)
+\tfrac{7}{15} \sin^4 (\theta /2)+\tfrac{328}{945} \sin^6 (\theta /2)\right)
\\
\phantom{e^{i \theta(\hat{\boldsymbol{n}}\cdot\boldsymbol{J})} = }{}
+\tfrac{1}{5!} (2i \hat{\boldsymbol{n}}\cdot\boldsymbol{J} \sin(\theta/2))^5 \cos(\theta/2)\left(1+\tfrac{4}
{3}\sin^2(\theta/2) +\tfrac{13} {9}\sin^4(\theta/2)\right)
\\
\phantom{e^{i \theta(\hat{\boldsymbol{n}}\cdot\boldsymbol{J})} = }{}
+\tfrac{1}{6!} {(2i \hat{\boldsymbol{n}}\cdot\boldsymbol{J} \sin(\theta/2))^6}\left(1+\sin^2 (\theta
/2)+\tfrac{13}{15} \sin^4 (\theta /2)\right)
\\
\phantom{e^{i \theta(\hat{\boldsymbol{n}}\cdot\boldsymbol{J})} = }{}
+\tfrac{1}{7!} (2i \hat{\boldsymbol{n}}\cdot\boldsymbol{J} \sin(\theta/2))^7 \cos(\theta/2)
\left(1+\tfrac{5}{3}\sin^2(\theta/2)\right)
\\
\phantom{e^{i \theta(\hat{\boldsymbol{n}}\cdot\boldsymbol{J})} = }{}
+\tfrac{1}{8!} {(2i \hat{\boldsymbol{n}}\cdot\boldsymbol{J} \sin(\theta/2))^8}\left(1+\tfrac{4}{3} \sin^2 (\theta /2)\right)
\\
\phantom{e^{i \theta(\hat{\boldsymbol{n}}\cdot\boldsymbol{J})} = }{}
+\tfrac{1}{9!} (2i \hat{\boldsymbol{n}}\cdot\boldsymbol{J} \sin(\theta/2))^9 \cos(\theta/2)
\\
\phantom{e^{i \theta(\hat{\boldsymbol{n}}\cdot\boldsymbol{J})} = }{}
+\tfrac{1}{10!} (2i \hat{\boldsymbol{n}}\cdot\boldsymbol{J} \sin(\theta/2))^{10},
\end{gather*}
etc.

The trace of each such exponential $(2j+1) \times (2j+1)$ matrix is the {\em character} of this spin~$j$ representation
of ${\rm SU}(2)$,~\cite[formulas (2.32)--(2.38)]{schwinger}, the Gegenbauer and also the 2nd kind Chebyshev polynomial,
$C^1_{2j}(\cos(\theta/2))=U_{2j} (\cos(\theta/2))=$ $\sin((2j+1)\theta/2)/\sin(\theta/2)$.
One might note that the above expansions may be expressed as linear combinations of Chebyshev polynomials, as well.
Thus, the coef\/f\/icient of~$(\hat{\boldsymbol{n}}\cdot \boldsymbol{J})^{k}$, for integer (semi-integer)~$j$
is, respectively, $\sum\limits_{\text{even (odd)}~n\leq2j}a_{n}T_{n}(\cos(\theta/2))$ for~$k$
even, and $\sum\limits_{\text{even (odd)}~n\leq2j}b_{n}V_{n}(\cos(\theta/2))$ for~$k$ odd,
where $T_{n}(\cos(\theta/2)) =\cos(n\theta/2)$ (Chebyshev 1st kind) and
$V_{n}(\cos(\theta/2)) =\sin(n\theta/2) =\sin(\theta/2)U_{n}(\cos(\theta/2))$ (Chebyshev 2nd kind).

It is evident by taking~$k$ derivatives of~\eqref{sum} with respect to~$\theta$ and evaluating at $\theta=0$ that the
$(i \hat{\boldsymbol{n}}\cdot\boldsymbol{J})^k$ term in the series is selected with unit coef\/f\/icient~-- all other terms must vanish.
It is also necessary, as manifest above, that the last term is of order $2j$.

For a~given~$j$, the matrix coef\/f\/icients of the terms of $\sin^{2j}(\theta/2)$ and $\cos(\theta/2)
\sin^{2j-1}(\theta/2)$ in the expansion, polynomials of order $(\hat{\boldsymbol{n}}\cdot\boldsymbol{J})^{2j}$ and
$(\hat{\boldsymbol{n}}\cdot\boldsymbol{J})^{2j-1}$ respectively, must vanish for all eigenvalues of $J_3$ except the
extremal ones, $\pm j$.
That is, they are proportional to the cha\-rac\-teristic polynomial of $J_3[j - 1]$ and hence of $\hat{\boldsymbol{n}}\cdot\boldsymbol{J} [j - 1]$.
Constraining these coef\/f\/icients to zero, which amounts to projecting out these extremal eigenvalues, necessarily reduces
the expansion of $e^{i \theta(\hat{\boldsymbol{n}}\cdot\boldsymbol{J}[j])}$ to that of $e^{i
\theta(\hat{\boldsymbol{n}}\cdot\boldsymbol{J}[j-1])} $.
The reverse procedure, recursive construction of $\exp({i \theta(\hat{\boldsymbol{n}}\cdot\boldsymbol{J}[j])})$ out of
$\exp ({i \theta(\hat{\boldsymbol{n}}\cdot\boldsymbol{J}[j - 1])})$ likewise follows.

For example, the spin $j=3/2$ expression leads to the spin $j=5/2$,
\begin{gather*}
e^{i \theta(\hat{\boldsymbol{n}}\cdot\boldsymbol{J}[5/2])} =
e^{i \theta(\hat{\boldsymbol{n}}\cdot\boldsymbol{J}[3/2])}\big\vert_{[5/2]}
+\big[\tfrac{1}{5!} (2i \sin(\theta/2))^5 \hat{\boldsymbol{n}}\cdot\boldsymbol{J}[5/2] + \tfrac{1}{4!} \cos(\theta/2)(2i \sin(\theta/2))^4 \big]
\\
\phantom{e^{i \theta(\hat{\boldsymbol{n}}\cdot\boldsymbol{J}[5/2])} =}{}
\times \bigl [\bigl((\hat{\boldsymbol{n}}\cdot\boldsymbol{J}[5/2])^2 - \big(\tfrac{3}{2}\big)^2\bigr) \bigl
((\hat{\boldsymbol{n}}\cdot\boldsymbol{J}[5/2])^2 - \big(\tfrac{1}{2}\big)^2\bigr) \bigr],
\end{gather*}
where the f\/irst term on the right hand side means the polynomial of~\eqref{quartet} evaluated for the spin $j=5/2$,
$6\times6$ matrices, instead of the $4\times4$ ones.

All terms, for integral~$j$ are actually periodic functions~$\theta$, i.e.~they can be recast in trigonometric functions
of~$\theta$, and are thus periodic in $2\pi$.
By contrast, for half-integral spin~$j$, all coef\/f\/icients cannot (they are only trigonometric functions of $\theta/2$),
and and thus have period $4\pi$, instead.

Note the alternating binary parities~$\epsilon$, even (0), or odd (1~-- which inserts a~factor $\cos(\theta/2)$ in the
coef\/f\/icients and ef\/fectively in the denominator of the truncated series to be discussed), interleave for a~given~$j$;
and their location shifts by one for a~given~$k$, going from integral to half-integral spins.

Thus, the odd $\epsilon =1$ series for $k=0$, so then for half-integral spins, truncated to $\lfloor j\rfloor=j-1/2$
terms past its leading term~1 is
\begin{gather*}
c_0(\epsilon =1) /\cos(\theta/2) = 1+\tfrac{1}{2} x+ \tfrac{3}{8} x^2+ \tfrac{5}{16} x^3 +\cdots + \binom{2j-1}{j-1 /
2}\left(\frac{x}{4}\right)^{j-1/2}
\\
\phantom{c_0(\epsilon =1) /\cos(\theta/2)}{}
 =\frac{1}{\sqrt{\pi}} \sum\limits_{n=0}^{j-\tfrac{1}{2}} \frac{\Gamma\left(n+\frac{1}{2}\right)}
{\Gamma\left(n+1\right)}x^{n}.
\end{gather*}

Note the contrast to the likewise odd $\epsilon =1$ series for $k=1$, thus integral spins, truncated to $\lfloor j-1/2
\rfloor=j-1$ terms past its leading term,
\begin{gather*}
c_1(\epsilon =1) /\cos(\theta/2) = 1+\tfrac{2}{3} x+ \tfrac{8}{15} x^2+ \tfrac{16}{35} x^3+\cdots + \frac{(j-1)!(j-1)!}{(2j-1)!}
(4x)^{j-1}
\\
\phantom{c_1(\epsilon =1) /\cos(\theta/2)}
= \sqrt{\pi} \sum\limits_{n=0}^{j-1} \frac{\Gamma\left(n+1\right)} {\left(2n+1\right)
\Gamma\left(n+\tfrac{1}{2}\right)}x^{n}.
\end{gather*}

When the large~$j$ limit is considered, the resulting expression would be expected to resemble the expansion of a~scalar
exponential, as the Cayley--Hamilton theorem applied to higher-order terms provides dwindling corrections~-- provided the
requisite periodicities are respected! The leading coef\/f\/icient (to the identity) of the expansion, $c_0$, is always
just~1, for even~$\epsilon$, so then for all integral spins, large and small.

In striking contrast, for odd~$\epsilon$, large half-integral spins~$j$, the leading term tends to
\begin{gather*}
I_{2j+1}\textrm{sgn} \bigl (\cos (\theta/2)\bigr),
\end{gather*}
a~square waveform with the required periodicity of $4\pi$.
It agrees with the integral spin in $[-\pi, \pi]$, but f\/lips sign outside this interval.
(See the f\/irst graph in the appendix.)

Similarly, the second term in the expansion linear in the Lie algebra generators, for large integral spins has odd
bimodal parity~$\epsilon$ and tends to
\begin{gather*}
i \hat{\boldsymbol{n}}\cdot\boldsymbol{J}\left(\theta - 2\pi \lfloor \tfrac{\theta}{2\pi} - \tfrac{1}{2} \rfloor\right),
\end{gather*}
a~sawtooth forced to maintain periodicity in $2\pi$.
(See the second graph in the appendix.)

By contrast, for large half-integral spins, even bimodal parity, the limiting triangular waveform can be more symmetric,
as the slope of the linear function must reverse at the boundary of $[-\pi, \pi]$,
\begin{gather*}
i \hat{\boldsymbol{n}}\cdot\boldsymbol{J}\textrm{sgn} \bigl (\cos (\theta /2)\bigr)
\bigl(\theta - 2\pi \lfloor\tfrac{\theta}{2\pi} - \tfrac{1}{2} \rfloor\bigr).
\end{gather*}
Thus, for large~$j$, all coef\/f\/icients tend to trigonometric series representations~\cite{pinsky,zygmund} as discussed in
the last section and illustrated in the appendix.

\section{Equivalent, top-down parameterization\\ of the series of coef\/f\/i\-cients}\label{section4}

The expansion introduced,~\eqref{sum}, may be re-expressed in a~descending powers' sum, with the powers of sine now
incorporated in the coef\/f\/icients, $C_k [j]=c_k (i\sin(\theta/2))^k/k!$, for convenience,
\begin{gather*}
e^{i \theta(\hat{\boldsymbol{n}}\cdot\boldsymbol{J} [j])} = \sum\limits_{m=0}^{2j} C_{2j-m} [j]
\big(2\hat{\boldsymbol{n}}\cdot\boldsymbol{J} [j]\big)^{2j-m}.
\end{gather*}
The representation index $[j]$ is also displayed explicitly here since, in the next section, dif\/ferent spins can enter
in the same formula; normally this argument may be suppressed when working within a~representation, as has been the case
so far.

The top-down coef\/f\/icients are analytic functions of~$j$ and apply to either integer or semi-integer spin,
\begin{gather}
 C_{2j} [j] =\tfrac{1}{(2j)!}\left(i \sin (\theta/2)\right)^{2j},
\label{last}
\\
C_{2j-1} [j]  =\tfrac{1}{(2j-1) !}\left(\cos(\theta/2)\right)\left(i \sin(\theta/2)\right)^{2j-1},
\label{penultimate}
\\
C_{2j-2} [j]  =\tfrac{1}{\left(2j-2\right) !}\left(i \sin(\theta/2)\right)^{2j-2}\left(1+\tfrac{1}{3}\left(j-1\right)
\sin^2 (\theta/2)\right),
\nonumber
\\
C_{2j-3} [j]  =\tfrac{1}{(2j-3) !}\left(\cos (\theta/2)\right)\left(i \sin (\theta/2)
\right)^{2j-3}\left(1+\tfrac{1}{3}j \sin^{2}(\theta/2)\right),
\nonumber
\\
C_{2j-4} [j] =\tfrac{1}{(2j-4) !}\left(i \sin(\theta/2)\right)^{2j-4}
\nonumber
\\
\phantom{C_{2j-4} [j]=}{}
 \times\left(1+\tfrac{1}{3}(j-2) \sin^{2}(\theta/2)+\tfrac{1}{90}\left(5j+1\right)\left(j-2\right)
\sin^{4}(\theta/2)\right),
\nonumber
\\
C_{2j-5} [j] =\tfrac{1}{(2j-5) !}\left(\cos(\theta/2)\right)\left(i \sin(\theta/2)\right)^{2j-5}
\nonumber
\\
\phantom{C_{2j-5} [j]=}{}
 \times\left(1+\tfrac{1} {3}\left(j-1\right) \sin^{2}(\theta/2) +\tfrac{1}{90}\left(5j+1\right)
j\sin^{4}(\theta/2)\right),
\nonumber
\end{gather}
etc.

In general then, for any integer~$n$, so long as $n\leq j$,
\begin{gather}
C_{2j{-}2n{+}1} [j] =\frac{\left(i \sin\left(\theta/2\right)\right)^{2j{-}2n{+}1} \cos(\theta/2)}{(2j-2n+1)
!}\underset{n-1}{\operatorname*{Trunc}} \! \left[ \frac {1}{\sqrt{1\!-\!x}}
\left(\frac{\arcsin\!\sqrt{x}}{\sqrt{x}}\right)^{2j{-}2n{+}1} \!\right\vert_{x=\sin^{2}(\theta/2)}\!\!\!\!,     \!\!\!\!\!\!\!\!\!\!\!\!\!
\label{Odd}
\\
C_{2j-2n} [j] =\frac{\left(i\sin (\theta/2)\right)^{2j-2n}}{\left(2j-2n\right) !} \underset{n}{\operatorname*{Trunc}}
\left[\left(\frac{\arcsin\sqrt{x}}{\sqrt{x}}\right)^{2j-2n}\right\vert_{x=\sin^{2}(\theta/2)},
\label{Even}
\end{gather}
as $\epsilon(2n-1)=1$, and $\epsilon (2n)=0$, respectively.
Of course, $1=\underset{n}{\operatorname*{Trunc}}(1)$ for any $n\geq0$.

Note from these expressions that half of the~$C$s for a~given~$j$ are related by a~simple derivative recursive condition
between contiguous coef\/f\/icients,{\samepage
\begin{gather*}
i C_{2(j-n) -1} [j] =2\frac{d}{d\theta}C_{2(j-n)} [j]
\qquad
\text{for}
\quad
n\geq0,
\end{gather*}
that is, derivation of the $\epsilon=0$ terms yields the contiguous $\epsilon=1$ terms.}

The sequence of coef\/f\/icients terminates at $C_{0} [j]$, of course, which, on the one hand, for integer~$j$ requires
$n=j$ in~\eqref{Even}, giving $C_{0} [n]=1$.

But, on the other hand, for half integer~$j$, it requires $n=j+1/2$ in~\eqref{Odd}, giving
\begin{gather*}
C_{0} [\tfrac{2n-1}{2}] = \cos(\theta/2)\left.
\operatorname*{Trunc}_{\ n-1}\left(\frac{1}{\sqrt{1-x}}\right) \right\vert_{x=\sin^{2} (\theta/2)}. 
\end{gather*}
Moreover, for half-integral spins, derivation of this leading, $\epsilon=1$ term, yields the last, $\epsilon=0$ term,
\begin{gather*}
2i\frac{d}{d\theta}C_{0} [j] = C_{2j}[j]\left(\frac{(2j!)}{(j-1/2)!2^{j-1/2}}\right)^2 (-1)^{j-1/2}.
\end{gather*}

This is the only distinction between integer and semi-integer spins in this analytic top-down approach, although these two $C_{0}$s 
can be compacted to a~single expression upon using the familiar discrete \emph{Bose/Fermi index}, or spin
discriminant, a~binary variable,
\begin{gather*}
\epsilon (j ) =\big(1- (-1 )^{2j}\big) /2=2(j - \lfloor j \rfloor) =
\begin{cases}
0 & \text{if} \  j \ \text{is integer},
\\
1 & \text{if} \  j \ \text{is half-integer}.
\end{cases}
\end{gather*}

By use of the same discriminant, the coef\/f\/icients can be written in a~common form for either an even or an odd number of
steps down from $2j$.
 Thus for any descending order $m\in[0,1,\dots,2j] $, we have $\epsilon(m/2) =1$ if~$m$ is an odd
integer and $\epsilon(m/2) =0$ if~$m$ is an even integer, so that
\begin{gather*}
C_{2j{-}m} [j] =\frac{\bigl (\cos(\theta/2) \bigr)^{\epsilon (m/2)} \bigl (i \sin(\theta/2) \bigr)^{2j{-}m}}{(2j-m)!}
\underset{\left\lfloor
\frac{m}{2}\right\rfloor}{\operatorname*{Trunc}}
\left[\left(\frac{1}{\sqrt{1-x}}\right)^{\epsilon (m/2 )}\left(\frac{\arcsin\sqrt{x}}{\sqrt{x}}\right)^{2j{-}m}\right]\!,
\end{gather*}
where $x\equiv\sin^{2}(\theta/2)$ and $\left\lfloor \dots\right\rfloor$ is the f\/loor function, with $n=\left\lfloor
\frac{2n}{2}\right\rfloor =\left\lfloor \frac{2n+1} {2}\right\rfloor $.
The interchanged roles of ascending and descending indices is evident upon comparison with the equivalent
formula~\eqref{sum}.


\section{Generating functions}\label{section5}

In some contrast to the previous section, derivatives of $\epsilon=1$ coef\/f\/icients are related to other coef\/f\/icients,
but, in general, of {\em lower spins}~$j$ and of both $\epsilon=0$ and $\epsilon=1$.

For instance, for integer~$j$ and integer~$n$, and for $C_0[0]\equiv 1$, $C_1[0]\equiv 0$,
\begin{gather*}
\left(2\tfrac{d}{d\theta}+\tan(\theta/2)\right)C_{2n+1}[j]
=\sum\limits_{m=0}^{j-n-1}\Big(i\cos^2 (\theta/2) \sin^{2m} (\theta/2) C_{2n}[j-m-1]
\\
\phantom{\left(2\tfrac{d}{d\theta}+\tan(\theta/2)\right)C_{2n+1}[j]=}{}
+\cos (\theta/2) \sin^{2m+1}(\theta/2) C_{2n+1}[j-m-1]\Big).
\end{gather*}
For half-integer~$j$, likewise,
\begin{gather*}
\left(2\tfrac{d}{d\theta}+\tan(\theta/2)\right)C_{2n}[j]
=\sum\limits_{m=0}^{j-n-3/2}\Big(i \cos^2 (\theta/2) \sin^{2m} (\theta/2) C_{2n-1}[j-m-1]
\\
\phantom{\left(2\tfrac{d}{d\theta}+\tan(\theta/2)\right)C_{2n}[j]=}{}
+ \cos (\theta/2)\sin^{2m+1} (\theta/2) C_{2n}[j-m-1]\Big).
\end{gather*}

The top-down coef\/f\/icients~\eqref{Odd}, \eqref{Even}
may be encoded systematically in a~pair of {\em Generating functions}
linking {\em all} dif\/ferent spins~$j$, and reliant on the incomplete Gamma function,
\begin{gather*}
\Gamma\left(n+1,z\right) \equiv\int_{z}^{\infty}e^{-t}t^{n} dt,
\qquad
\Gamma\left(n+1,0\right) =\Gamma\left(n+1\right) =n!,
\end{gather*}
albeit with operator arguments.

Letting $j=m/2$ and summing over $m=0,1,2,\dots,\infty$ to cover {\em all spins}, both integer and half integer, the
master generating functions are
\begin{gather*}
G_{2n}(t,x) \equiv\sum\limits_{m=0}^{\infty} t^{m}C_{m-2n} [m/2] =\sum\limits_{m=2n}^{\infty}t^{m}C_{m-2n}[m/2]
\\
\phantom{G_{2n}(t,x)}
=t^{2n}\left[\sum\limits_{r=0}^{n}\frac{x^{r}}{r!}\frac{d^{r}}{dy^{r}}\exp\left(\frac{t\sqrt{x}\arcsin\sqrt{y}}{\sqrt{y}}\right)
\right\vert_{y=0}
\\
\phantom{G_{2n}(t,x)}
=\frac{1}{n!}t^{2n}\left[\Gamma\left(n+1,x\frac{d}{dy}\right)
\exp\left(\frac{t\sqrt{x}\arcsin\sqrt{y+x}}{\sqrt{y+x}}\right) \right\vert_{y=0}
\end{gather*}
and
\begin{gather*}
 G_{2n+1}(t,x) \equiv\sum\limits_{m=0}^{\infty}t^{m}C_{m-2n+1} [m/2]
=\sum\limits_{m=2n-1}^{\infty}t^{m}C_{m-2n+1} [m/2]
\\
\phantom{G_{2n+1}(t,x)}
=t^{2n-1}\sqrt{1-x}\left[\sum\limits_{r=0}^{n-1}\frac{x^{r}}{r!}\frac{d^{r}}{dy^{r}}\left(\frac{1}{\sqrt{1-y}}
\exp\left(\frac{t\sqrt{x}\arcsin\sqrt{y}}{\sqrt{y}}\right)\right) \right\vert_{y=0}
\\
\phantom{G_{2n+1} (t,x)}
 =\frac{1}{n!}t^{2n{-}1}\sqrt{1\!-\!x}\left[\Gamma\!\left(n+1,x\frac{d} {dy}\right)
\left(\frac{1}{\sqrt{1\!-\!x\!-\!y}}\exp\left(\frac{t\sqrt{x}\arcsin\sqrt{y\!+\!x}}{\sqrt{y\!+\!x}}\right)\right) \right\vert_{y=0}\!.
\end{gather*}

Thus, for instance,
\begin{gather*}
G_{0} =\sum\limits_{m=0}^{\infty}t^{m}C_{m} [m/2] =\sum\limits_{m=0}^{\infty}\frac{t^{m}}{m!} \bigl (i \sin (\theta/2)
\bigr)^{m} = e^{it\sin (\theta/2)},
\end{gather*}
yields~\eqref{last} for each $j=m/2$;
\begin{gather*}
G_{1}  =\sum\limits_{m=0}^{\infty}t^{m}C_{m-1} [m/2] =\sum\limits_{m=1}^{\infty} t^m C_{m-1} [m/2]
\\
\phantom{G_{1}}
 =\sum\limits_{m=1}^{\infty}\frac{t^{m}}{\left(m-1\right) !}\left(\cos (\theta/2)\right)\left(i \sin
(\theta/2)\right)^{m-1}=\left(\cos (\theta/2)\right) te^{it\sin (\theta/2)},
\end{gather*}
yields~\eqref{penultimate} for each $j=m/2$;
\begin{gather*}
G_{2} =\sum\limits_{m=0}^{\infty}t^{m}C_{m-2} [m/2] =\sum\limits_{m=2}^{\infty} t^m C_{m-2} [m/2]
\\
\phantom{G_{2}}
 = \sum\limits_{m=2}^{\infty}\frac{t^{m}}{\left(m-2\right) !}\left(i \sin (\theta/2)\right)^{m-2} \Bigl (1+\tfrac{1}{3}
\big(\tfrac{m}{2} -1\big) \sin^{2} (\theta/2)\Bigr)
\\
\phantom{G_{2}}
 =t^{2}\operatorname{KummerM}\left(\frac{\sin^{2} (\theta/2) +6}{\sin^{2} (\theta/2)},
\frac{6}{\sin^{2}(\theta/2)},it\sin (\theta/2)\right),
\end{gather*}
etc., involving hypergeometric functions.

\section{Specif\/ication of the series}\label{section6}

Calculation of the rotation matrices as spin polynomials can be ef\/f\/iciently carried out in specif\/ic cases using either
Lagrange--Sylvester expansions, or the Cayley--Hamilton theorem.
These two methods are tied together by Vandermonde matrix algebra~\cite{atiyah}.
The expressions for general spin $j$ given in the f\/irst section of this paper can be established using these methods.

Consider functions of an $N\times N$ diagonalizable matrix $\mathbb{M}$ with non-degenerate eigenvalues $\lambda_{i}$,
$i=1,\dots,N$.
 On the span of the eigenvectors, there is an obviously correct Lagrange formula, as extended to matrices by Sylvester,
\begin{gather}
f\left(\mathbb{M}\right) =\sum\limits_{i=1}^{N}f\left(\lambda_{i}\right) \mathbb{P}_{i},
\label{39}
\end{gather}
where the projection operators~-- the so-called Frobenius covariants~-- are given by products,
\begin{gather}
\mathbb{P}_{i}= {\displaystyle\prod\limits_{\substack{j=1\\j\neq i}}^{N}} \frac{\mathbb{M}-\lambda_{j}}{\lambda_{i}-\lambda_{j}}.
\label{Proj}
\end{gather}
From expanding each such product it is evident that any $f(\mathbb{M})$ reduces to a~polynomial of order $N-1$ in powers
of $\mathbb{M}$,
\begin{gather}
f\left(\mathbb{M}\right) =\sum\limits_{m=0}^{N-1}C_{m}\left[f\right] \mathbb{M}^{m},
\label{fPoly}
\end{gather}
and the function-dependent coef\/f\/icients can be expressed in terms of the eigenvalues of $\mathbb{M}$\ by expanding the 
projection operators~\eqref{Proj} as polynomials in $\mathbb{M}$.

The condition that no two eigenvalues be the same in~\eqref{Proj} can be ef\/f\/iciently implemented using Vandermonde
matrix methods.
 From the polynomial~\eqref{fPoly} acting on eigenvector $\left\vert \lambda_{k}\right\rangle$ we obtain
\begin{gather*}
f\left(\lambda_{k}\right) =\sum\limits_{m=0}^{N-1}C_{m}\left[f\right] \left(\lambda_{k}\right)^{m}.
\end{gather*}
The action on the full set of~$N$ eigenvectors therefore gives an $N\times N$ matrix equation,
\begin{gather}
\left(
\begin{matrix}
f(\lambda_{1})
\\
f(\lambda_{2})
\\
\vdots
\\
f(\lambda_{N})
\end{matrix}
\right) =V\left[\mathbb{M}\right]\left(
\begin{matrix}
C_{0}[f]
\\
C_{1}[f]
\\
\vdots
\\
C_{N-1}[f]
\end{matrix}
\right),
\label{f=VC}
\end{gather}
where the $N\times N$ Vandermonde matrix of $\mathbb{M}$ eigenvalues is
\begin{gather*}
V\left[\mathbb{M}\right] =\left(
\begin{matrix}
1 & \lambda_{1} & \lambda_{1}^{2} & \dots & \lambda_{1}^{N-1}
\\
1 & \lambda_{2} & \lambda_{2}^{2} & \dots & \lambda_{2}^{N-1}
\\
\vdots & \vdots & \vdots & \ddots & \vdots
\\
1 & \lambda_{N} & \lambda_{N}^{2} & \dots & \lambda_{N}^{N-1}
\end{matrix}
\right).
\end{gather*}

As is well-known, for nondegenerate eigenvalues, $V$ is nonsingular~\cite{macon}.
Thus, one may obtain the coef\/f\/icients $C_{k}[f]$ in the expansion~\eqref{fPoly} by merely inverting~$V$
in~\eqref{f=VC}, into the inverse relation,
\begin{gather}
\left(
\begin{matrix}
C_{0}[f]
\\
C_{1}[f]
\\
\vdots
\\
C_{N-1}[f]
\end{matrix}
\right) =V^{-1}\left[\mathbb{M}\right]\left(
\begin{matrix}
f(\lambda_{1})
\\
f(\lambda_{2})
\\
\vdots
\\
f(\lambda_{N})
\end{matrix}
\right),
\label{C=invVf}
\end{gather}
although one should note that, by our conventions, the row index~$m$ runs from $0$ to $N-1$, while the column index~$i$
appears more familiar, running from $1$ to~$N$.


For the problem at hand, we are interested in the case where the matrix is
$\mathbb{M}=2\hat{\boldsymbol{n}}\cdot\boldsymbol{J}$.
 Thus for spin~$j$ we have a~matrix $\mathbb{M}$ whose dimension is $N=2j+1$ and whose eigenvalues are
$\lambda_{k}=2(j+1-k)$, ordered so that $\left\{\lambda_{1},\lambda_{2},\dots,\lambda_{N}\right\}
=\left\{2j,2j-2,\dots,-2j\right\}$.
 The Vandermonde matrix for spin~$j$ is then $(2j+1) \times(2j+1)$ as given~by
\begin{gather*}
V[j] =\left(
\begin{matrix}
1 & 2j & (2j)^{2} & \dots & (2j)^{2j}
\\
1 & 2j-2 & (2j-2)^{2} & \dots & (2j-2)^{2j}
\\
\vdots & \vdots & \vdots & \ddots & \vdots
\\
1 & -2j & (-2j)^{2} & \dots & (-2j)^{2j}
\end{matrix}
\right).
\end{gather*}
Note here that $\det V[j] \neq0$, hence $V^{-1}[j]$ exists.
 In fact,
\begin{gather*}
\det V[j] =(-1)^{\left\lfloor \frac{2j+1}
{2}\right\rfloor}\prod\limits_{\text{prime}~p\leq2j}p^{m_{p}(j)},
\end{gather*}
where $m_{p}(j)$ is the multiplicity of the prime~$p$ that occurs in the factorization of $\det V[j]$.
Explicitly,
\begin{gather*}
m_{2}(j)  =\sum\limits_{k=1}^{2j}\sum\limits_{m=0}^{\left\lfloor \ln k/\ln2\right\rfloor}\left\lfloor
\frac{k}{2^{m}}\right\rfloor,
\qquad
\text{and}
\\
m_{p}(j) +j\left(2j+1\right)  =\sum\limits_{k=1}^{2j}\sum\limits_{m=0}^{\left\lfloor \ln k/\ln
p\right\rfloor}\left\lfloor \frac{k}{p^{m}}\right\rfloor, 
\qquad
\text{for}
\quad
p\geq3.
\end{gather*}
Note that a~given prime $p>2$ f\/irst appears in the factorization of the determinant for fermionic spin $j=p/2$, and
subsequently appears as a~factor in $\det V[j]$ for \emph{all} higher spins.

Returning to the problem at hand, we are interested in the exponential function, $\exp(\alpha\mathbb{M})$,
so by~\eqref{fPoly},
\begin{gather}
\exp\left(2\alpha\hat{\boldsymbol{n}}\cdot\boldsymbol{J}\right) =\sum\limits_{m=0}^{2j}C_{m}[j]
\left(2\hat{\boldsymbol{n}} \cdot\boldsymbol{J}\right)^{m},
\label{ExpAsPoly}
\end{gather}
where each~$j$ dependent coef\/f\/icient $C_{m}[j]$ is also implicitly a~function of~$\alpha$, to be determined.
 Of course, for rotations $\alpha=i\theta/2$.
 These functions are obtained by~\eqref{C=invVf}) specialized to this case, namely,
\begin{gather}
\left(
\begin{matrix}
C_{0}[j]
\\
C_{1}[j]
\\
\vdots
\\
C_{2j}[j]
\end{matrix}
\right) =V^{-1}[j]\left(
\begin{matrix}
e^{2j\alpha}
\\
e^{(2j-2) \alpha}
\\
\vdots
\\
e^{-2j\alpha}
\end{matrix}
\right).
\label{C=VinvExps}
\end{gather}
Thus the $C_{k}$s  
are linear combinations of the $2j+1$ exponentials shown.
 It remains only to compute the inverse of the Vandermonde matrix for spin~$j$, and to express the results for the
$C_{k}$s in a~compact form. 

One immediate result involves $C_{0}$ for integer spins:  Given our ordering of the eigenvalues, the f\/irst row of
$V^{-1}[j~\text{integer}]$ consists entirely of zeroes, except for a~$1$ in the middle column (i.e.~the $(j+1)$st)
corresponding to the obvious fact that the middle row (again the $(j+1) $st) of
$V[j~\text{integer}]$ has a~$1$ in the f\/irst column and zeroes in all subsequent columns (since $0^{k}=0$ for all positive~$k$).
 But the middle entry in the column of exponentials on the r.h.s.\  of~\eqref{C=VinvExps} is just $e^{0}=1$.
 Therefore $C_{0}=1$ for all integer~$j$.

All the coef\/f\/icients $C_{k}$ satisfy f\/irst-order dif\/ferential relations.
 Half of these relations are simply of the form $C_{n-1}=dC_{n}/d\alpha$ while the other half are more involved.
Above, we presented an interesting form for the more elaborate cases, where a~f\/irst-order equation
\emph{mixed spin $j$ coefficients with all lower spins} of the same type, either all integer or all semi-integer.
 Here, we take a~dif\/ferent approach, so that dif\/ferent spin coef\/f\/icients are not mixed ab initio.
 Nonetheless, in our labeling of the coef\/f\/icients as $C_{m}[j] $, we have chosen to keep track of the
spin~$j$ in addition to the power of $\hat{\boldsymbol{n}}\cdot\boldsymbol{J}$, as in the previous section, because the
form of some key results can be used to obtain recursion relations relating dif\/ferent spins.

First consider the simplest results which apply to half the coef\/f\/icients.
 For clarity, we give separately the results for the integer and semi-integer cases.
 For integer~$j$,
\begin{gather}
\frac{d}{d\alpha}C_{2k}[j] =C_{2k-1}[j],
\label{Coef 2k Integer j}
\end{gather}
where, for spin zero, $C_{0}[0] \equiv1$, and $C_{1}[0] \equiv0$.
 For semi-integer~$j$,
\begin{gather}
\frac{d}{d\alpha}C_{2k+1}[j] =C_{2k}[j].
\label{Coef 2k+1 Semi-Integer j}
\end{gather}

We now describe an approach which applies to all other cases just as well.
Dif\/ferentiating the coef\/f\/icients in the expansion of the exponential~\eqref{ExpAsPoly} gives on the one hand,
\begin{gather}
\frac{d}{d\alpha}\exp\left(2\alpha\hat{\boldsymbol{n}}\cdot\boldsymbol{J}\right)
=\sum\limits_{m=0}^{2j}\frac{d}{d\alpha}C_{m}[j] \left(2\hat{\boldsymbol{n}}\cdot\boldsymbol{J}\right)^{m},
\label{derivRHS}
\end{gather}
while, on the other hand, just from dif\/ferentiating the exponential itself, it follows that
\begin{gather}
\frac{d}{d\alpha}\exp\left(2\alpha\hat{\boldsymbol{n}}\cdot\boldsymbol{J}\right)
=2\hat{\boldsymbol{n}}\cdot\boldsymbol{J}\exp\left(2\alpha\hat{\boldsymbol{n}}\cdot\boldsymbol{J}\right)
\nonumber
\\
\phantom{\frac{d}{d\alpha}\exp\left(2\alpha\hat{\boldsymbol{n}}\cdot\boldsymbol{J}\right)}
 =\sum\limits_{m=1}^{2j}C_{m-1}[j] \left(2\hat{\boldsymbol{n}}\cdot\boldsymbol{J}\right)^{m}+C_{2j}[j]
\left(2\hat{\boldsymbol{n}}\cdot\boldsymbol{J}\right)^{2j+1}.
\label{derivLHS}
\end{gather}
We obtain f\/irst-order dif\/ferential relations for the coef\/f\/icients by equating the two expressions~\eqref{derivLHS}
and~\eqref{derivRHS}.
 To make further progress, it is necessary to resolve $(2\hat{\boldsymbol{n}} \cdot\boldsymbol{J})^{2j+1}$
as a~series in lower powers of $2\hat{\boldsymbol{n}}\cdot\boldsymbol{J}$.

This resolution can be achieved by applying the Cayley--Hamilton theorem to the spin $j>0$ matrices.
 The f\/inal result is straightforward to obtain:
\begin{gather}
\left(2\hat{\boldsymbol{n}}\cdot\boldsymbol{J}\right)^{2j+1}=\sum\limits_{m=0}^{2j}A_{m}[j]
\left(2\hat{\boldsymbol{n}} \cdot\boldsymbol{J}\right)^{m},
\label{2j+1 expansion}
\end{gather}
where the coef\/f\/icients are explicitly encoded in the following polynomial, valid for either integer or semi-integer
spin:
\begin{gather}
p_{j}\left(x\right) \equiv x^{2j+1}-\frac{1}{x^{2j+1}} {\displaystyle\prod\limits_{m=0}^{2j}}
\left(x^{2}-\left(\tfrac{1+(-1)^{m+2j}} {2}\right) m^{2}\right) =\sum\limits_{m=0}^{2j}A_{m} [j]x^{m}.
\label{coefficient poly}
\end{gather}
The resolution of $(2\hat{\boldsymbol{n}}\cdot\boldsymbol{J})^{2j+1}$ may also be reduced to a~Vandermonde
matrix equation when acting with~\eqref{2j+1 expansion} on the eigenvectors of $2\hat{\boldsymbol{n}}
\cdot\boldsymbol{J}$, and leads again to the results encoded in the $p_{j}(x)$ given in~\eqref{coefficient
poly}.

To connect with existing mathematical literature, it is worthwhile to note the coef\/f\/icients for semi-integer spins may
be appropriately called \emph{fermionic central factorial numbers}, i.e.~``$\operatorname*{fcfn}$'', but they are actually known in the literature as just \emph{central factorial
numbers}~\cite{CFN}\footnote{These are presented as triangular matrices in \url{http://oeis.org/A008956}.}.
Similarly, the $A_{m}[j]$ for integer spins may be called the \emph{bosonic central factorial numbers},
i.e.~``$\operatorname*{bcfn}$'', but they are known in the mathematics literature as the
\emph{scaled central factorial numbers}\footnote{These are presented as triangular matrices in \url{http://oeis.org/A182867}.}.
  By interlacing the rows and columns of the $\operatorname*{fcfn}$ and
$\operatorname*{bcfn}$ matrices, the coef\/f\/icients for both integer and semi-integer cases can be expressed in a~unif\/ied
manner in terms of a~single matrix, the result being ef\/fectively a~replication of~\eqref{coefficient poly}.

Combining~\eqref{derivLHS},~\eqref{derivRHS}, and~\eqref{2j+1 expansion} leads to
\begin{gather}
\frac{dC_{m}[j]}{d\alpha}=C_{m-1}[j] +C_{2j}[j] A_{m}[j]
\qquad
\text{for}
\quad
m=0,\dots,2j
\label{dC/dt C CA}
\end{gather}
with the convention $C_{-1}[j] =0$, which immediately shows two noteworthy features:  The simplicity of
$dC_{0}[j] /d\alpha$ and $dC_{1}[j] /d\alpha$ for \emph{any}~$j$.
 The coef\/f\/icients of the lowest and highest powers of $(2\hat{\boldsymbol{n}} \cdot\boldsymbol{J})$ are
linked by dif\/ferentiation.
 For example,
\begin{gather*}
\frac{d}{d\alpha}C_{0}[j] =(-1)^{\left\lfloor j\right\rfloor}\left[(2j)!!\right]^{2}C_{2j}[j]
\qquad
\text{for}
\quad
j
\
\text{semi-integer}.
\end{gather*}
Similarly, since $C_{0}[j~\text{integer}] =1$,
\begin{gather*}
\frac{dC_{1}[j]}{d\alpha}=1+(-1)^{j-1}\left[(2j) !!\right]^{2}C_{2j}[j]
\qquad
\text{for}
\quad
j
\
\text{integer}.
\end{gather*}

Symmetry under the ref\/lection map $\hat{\boldsymbol{n}}\cdot\boldsymbol{J}
\longmapsto-\hat{\boldsymbol{n}}\cdot\boldsymbol{J}$ implies only odd integer~$m$ contribute on the r.h.s.\  of~\eqref{2j+1
expansion} for integer~$j$, i.e.~$A_{2k}[j~\text{integer}] =0$, and only even integer~$m$ contribute for
half-integer~$j$, i.e.~$A_{2k+1}[j~\text{half-integer} ] =0$.
 This suf\/f\/ices to establish~\eqref{Coef 2k Integer j} and~\eqref{Coef 2k+1 Semi-Integer j}, since~\eqref{dC/dt C CA}
reduces to those simple results for $m=2k$ and integer~$j$, and for $m=2k+1$ and semi-integer~$j$, respectively.

The remaining cases are less trivial, as given~by
\begin{gather}
\frac{dC_{2k+1}[j]}{d\alpha}=C_{2k}[j] +\frac {1}{(2j)!}\left(\sinh\alpha\right)^{2j}A_{2k+1} [j]
\qquad
\text{for integer}
\
j,
\label{integer hierarchy}
\\
\frac{dC_{2k}[j]}{d\alpha}=C_{2k-1}[j] +\frac {1}{(2j)!}\left(\sinh\alpha\right)^{2j}A_{2k} [j]
\qquad
\text{for semi-integer}
\
j,
\label{semi-integer hierarchy}
\end{gather}
where we have used the fact that
\begin{gather*}
C_{2j}[j] =\frac{1}{(2j) !}\left(\sinh \alpha\right)^{2j},
\end{gather*}
manifest by inspection of the highest-order term in the Lagrange--Sylvester expansions,~\eqref{39}--\eqref{fPoly}.

Given the coef\/f\/icients $A_{m}[j]$ then, as provided by~\eqref{coefficient poly}, the coef\/f\/icients
$C_{m}[j]$ are computed successively from the lowest to the highest values of~$m$ by integration
of~\eqref{integer hierarchy} and~\eqref{semi-integer hierarchy}, or from highest to lowest values of~$m$, by carrying
out the dif\/ferentiations in those two equations.
 The results of this algorithm were given at the beginning of this paper.
(For a~distinctly dif\/ferent, albeit presumably equivalent algorithm, see~\cite{torruella}.)

So far in this section, only a~single spin~$j$ is involved in each relation.
Nevertheless, the explicit form for the polynomials in~\eqref{coefficient poly}) displays how the $p_{j}$ obey recursion
relations that relate dif\/ferent spins, namely,
\begin{gather*}
p_{j+1}(x) =(2j+2)^{2}x^{2j+1}+\big(x^{2}-(2j+2)^{2}\big)p_{j}(x).
\end{gather*}
This permits relating $A_{k}[j+1]$ and $A_{m}[j] $, and hence relating the derivatives of the
$C_{m}$s for dif\/ferent spins.
This, in turn, opens a~route to arriving at \emph{all} the f\/irst-order relations given in the previous section.

The explicit forms given here for the $C_{m}$s are, in fact, just the solutions of those f\/irst-order relations.
 In this way, a~rigorous proof of~\eqref{sum} may be obtained, as a~reader may readily verify~-- and illustrate for a~low~$j$.
Fortunately, it is not actually necessary to go through the straightforward but tedious details of such a~proof here:
Since this paper appeared on the arXiv,
a~simplif\/ied derivation, based on~\eqref{dC/dt C CA} and closed-form expressions
for $A_{m}[j]$ in terms of central factorial numbers has also appeared~\cite{TLCTVK}.

\section{Concluding remarks}\label{section7}

Rotation matrices for all representations~$j$ have been expressed as $2j$-order polynomials in the corresponding spin
matrices, equation~\eqref{sum}.
A~plethora of applications~\cite{torruella} of these elementary results can be easily imagined, across the entire
spectrum of quantum mechanical systems, including problems in atomic, condensed matter, nuclear, and elementary particle
physics~\cite{nelson,wdw}.

Such applications would range from the most straightforward, e.g.\
using the formula~\eqref{sum} to compute Wigner's~$d$ functions starting from a~specif\/ic representation of~$J_y$, to the
more esoteric, e.g., investigation of the large~$j$ asymptotics in various contrived matrix models.

Specif\/ically, consider
\begin{gather}
\det\left(\lambda \boldsymbol{1}- 2\hat{\boldsymbol{n}}\cdot\boldsymbol{J}\right) = \tfrac{2^{2j+1}}{\pi}\Gamma
\bigl(1+j+\tfrac{1}{2}\lambda \bigr) \Gamma \bigl(1+j-\tfrac{1}{2}\lambda \bigr) \sin \pi\left(\tfrac{1}{2}\lambda-j\right)
\nonumber
\\
\phantom{et\left(\lambda \boldsymbol{1}- 2\hat{\boldsymbol{n}}\cdot\boldsymbol{J}\right)}{}
 \underset{j\rightarrow \infty}{\sim} \frac{1}{\pi}2^{2j+1}\left(j!\right)^{2}\sin \pi\left(\tfrac{1}{2}\lambda
-j\right).
\label{CharacteristicAsymptotics}
\end{gather}
The two $\Gamma$ factors have poles that terminate the inf\/inite sequence of $\sin \pi (\frac{1}{2}\lambda
-j)$ zeroes both above and below, so that setting the r.h.s.\  of~\eqref{CharacteristicAsymptotics} to zero indeed
reproduces the f\/inite set of eigenvalues: $2j,2j-2,\dots,-2j+2,-2j$, as it should.

Alternatively, with the benef\/it of Stirling's formula,~\eqref{CharacteristicAsymptotics} becomes
\begin{gather*}
\prod\limits_{n=0}^{2j}\left(\lambda -2(j-n)\right) \underset{j\rightarrow \infty}{\sim} 2e^{-2j}(2j)^{2j+1}
\begin{cases}
(-1)^{j}\sin \pi (\frac{1}{2}\lambda) & \text{for}\  j\  \text{integer},
\\
(-1)^{j+\frac{1}{2}}\cos \pi (\frac{1}{2}\lambda) & \text{for}\  j\  \text{semi-integer},
\end{cases}
\end{gather*}
and thereby provides a~route to reach Euler's inf\/inite product representation of the sine and cosine functions.

In particular, as illustrated in Section~\ref{section3}, the large~$j$ limit warrants some additional mention here, since it is
intriguingly intuitive even in the completely quantum/noncommutative framework of this work.
In the limit $j\rightarrow \infty $, the coef\/f\/icients of the various powers,
$(2i\hat{\boldsymbol{n}}\cdot\boldsymbol{J})^n /n!$ in the expansion of the rotation group element become
quasi-classical: if $\hat{\boldsymbol{n}}\cdot\boldsymbol{J}$ were not a~matrix, but a~scalar instead, these
coef\/f\/icients in the exponential series would be just simple monomials, $(\theta/2)^n$.

Remarkably, the coef\/f\/icients presented in~\eqref{sum} reduce, in this limit, to elegant trigonometric
series~\cite{pinsky} for the periodicized monomials~-- exactly! --
\begin{gather*}
\lim_{j\rightarrow \infty} c_{k}\left(\theta\right) \sin^k ( \theta/2) =(-1)^{\left(1+\left\lfloor
\frac{\theta}{2\pi}-\frac{1}{2} \right\rfloor\right) \epsilon(j)}\left(\frac{1}{2}\left(\theta -2\pi -2\pi
\left\lfloor \frac{\theta}{2\pi}-\frac{1}{2} \right\rfloor\right)\right)^{k}.
\end{gather*}

Indeed, for f\/inite~$j$, the coef\/f\/icients are nothing but the Taylor polynomial truncations of these same trigonometric
series.
But even in the limit of large~$j$ there is always a~clear distinction between the integer and semi-integer spin cases,
as given by the fermionic sign f\/lip in this equation for $\pi <\left\vert \theta \right\vert <2\pi $.
This is illustrated in the appendix.

This appears as an elegant connection between the rotation group and Fourier analysis, moreover underlain by subtle
combinatorial identities involving the central factorial numbers, as we have explained in the previous section.
Thus, the results presented evoke an intriguing set of linkages between group theory, combinatorics, and analysis.

Extensions to ${\rm SO}(4)$ and ${\rm SU}_q(2)$ are direct.
In fact, the ${\rm SO}(4)$ case would be of some physical interest in superintegrable models involving extensions of the
Hermann--Bernoulli--Laplace--Hamilton--Gibbs--Runge--Lenz--Pauli vector to cases of arbitrary spin~\cite{Nikitin}.
Perhaps the results here can be of use in the analysis of such models.

\appendix
\section{Appendix.\ Periodicized monomials} \label{appendixA}

The remarks in the text about the behavior of the coef\/f\/icients in the large~$j$ limit are illustrated here for the
coef\/f\/icients of the lowest three powers of the generators with $-2\pi\leq\theta \leq2\pi$, for spins $j=137/2$ (shown in
blue) and $j=69$ (shown in purple).
 Even in the limit of large~$j$ there is always a~clear distinction between the integer and semi-integer spin cases, as
given by a~fermionic sign f\/lip for $\pi<\left\vert \theta\right\vert <2\pi$.

\begin{center}
\includegraphics{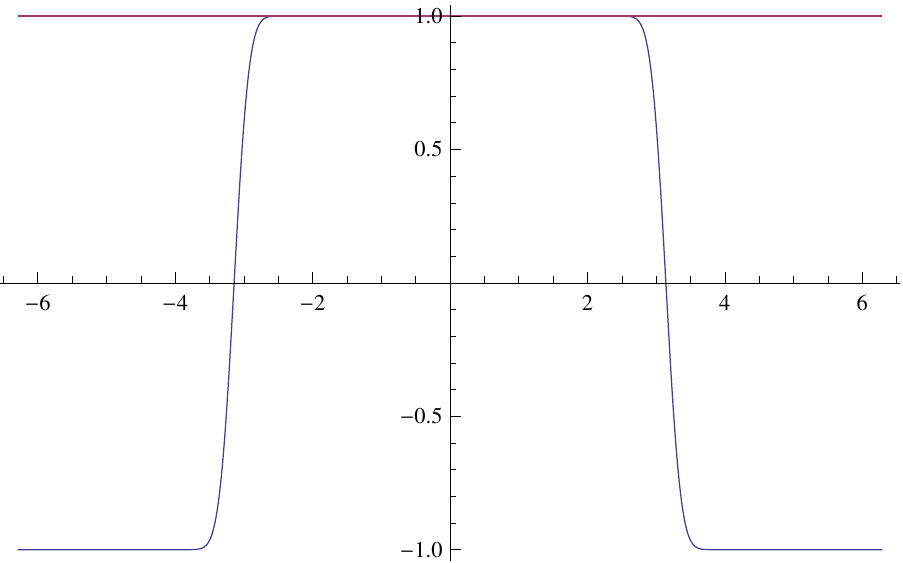}

$c_{0}$ versus~$\theta$
\end{center}

\begin{center}
\includegraphics{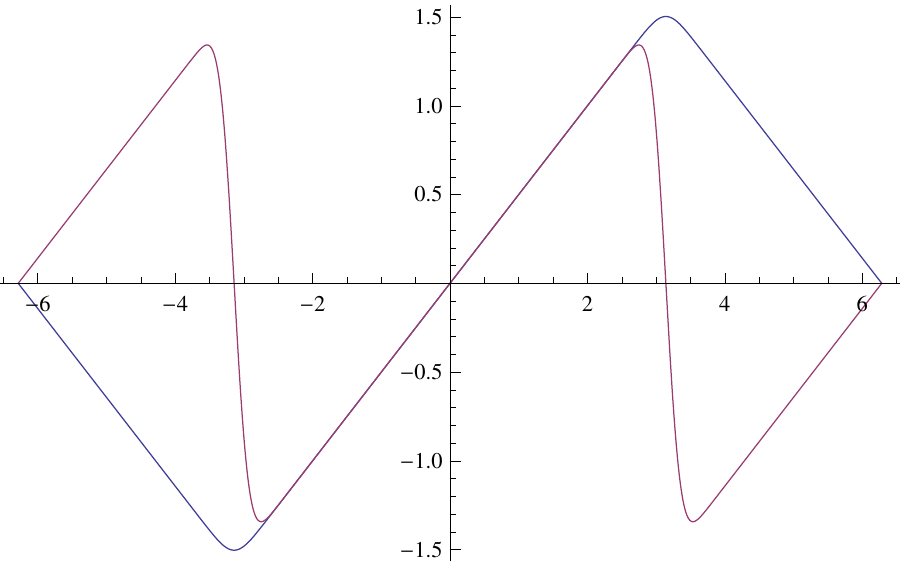}

$c_{1} \sin(\theta/2)$ versus~$\theta$
\end{center}

\begin{center}
\includegraphics{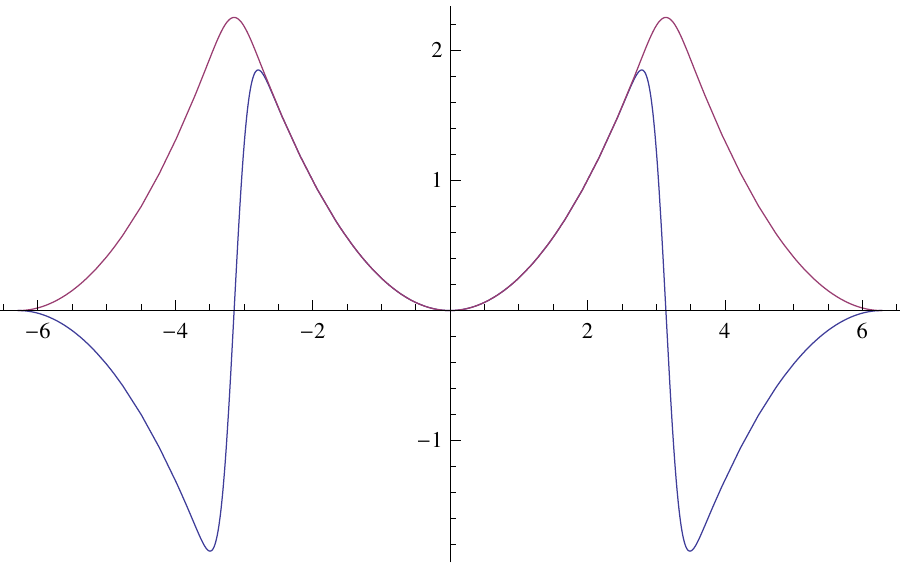}

$c_{2} \sin^2 (\theta/2)$ versus~$\theta$
\end{center}

\begin{center}
\includegraphics{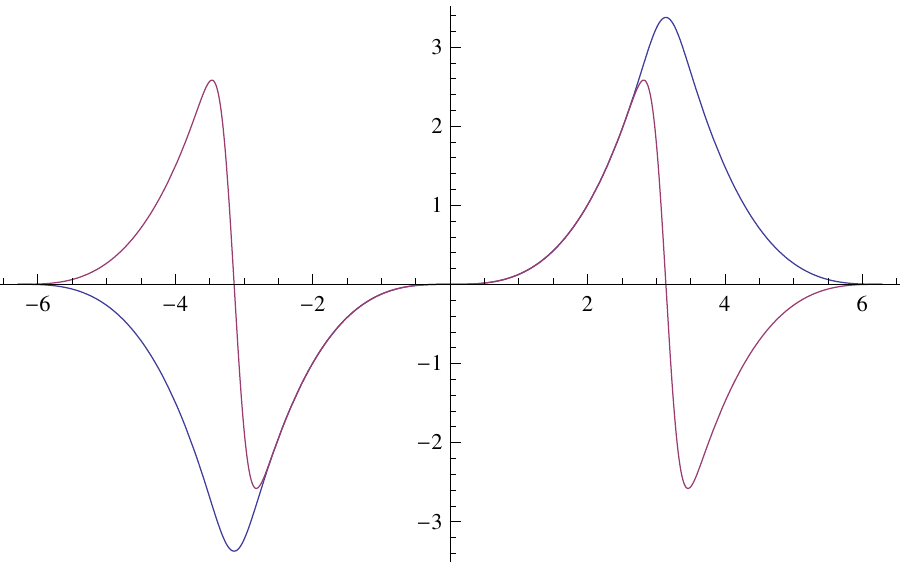}

$c_{3} \sin^3 (\theta/2)$ versus~$\theta$
\end{center}

These plots should be compared to the exact discontinuous functions obtained in the limit, as given by the last formula
in the text.
Note the striking absence of Gibbs--Wilbraham phenomena~\cite{pinsky} in these f\/igures.
Each term in the Taylor polynomials $\underset{n}{\operatorname{Trunc}}$ of~\eqref{trunco} is positive semi-def\/inite.

\subsection*{Acknowledgements}

\looseness=-1
The submitted manuscript has been created by UChicago Argonne, LLC, Operator of Argonne National Laboratory (Argonne).
Argonne, a~U.S.~Department of Energy Of\/f\/ice of Science laboratory, is operated under Contract No.~DE-AC02-06CH11357.
The U.S.~Government retains for itself, and others acting on its behalf, a~paid-up nonexclusive, irrevocable worldwide license in
said article to reproduce, prepare derivative works, distribute copies to the public, and perform publicly and display
publicly, by or on behalf of the Government.
It was also supported in part by NSF Award PHY-1214521.
TLC was also supported in part by a~University of Miami Cooper Fellowship.
S.~Dowker is thanked for bringing ref~\cite{torruella}, and whence~\cite{lehrer}, to our attention.
An anonymous referee is especially thanked for bringing~\cite{WW} and more importantly~\cite{wageningen} to our attention.

\pdfbookmark[1]{References}{ref}
\LastPageEnding

\end{document}